\documentstyle[preprint, aps, prc]{revtex}
\begin{document}
\draft
\bibliographystyle{prsty}
\title{Do Hadronic Charge Exchange Reactions Measure Electroweak $L=1$
Strength?}
\author{V. F. Dmitriev$^{1,2}$\thanks{Electronic address:
V.F.Dmitriev@inp.nsk.su}, V. Zelevinsky$^1$\thanks{Electronic
address: zelevinsky@nscl.msu.edu}, Sam
M.~Austin$^1$\thanks{Electronic address:  austin@nscl.msu.edu}}
\address{$^1$ National Superconducting
Cyclotron Laboratory and Department of Physics and Astronomy \\
Michigan State University, East Lansing, MI~48824\\  $^2$ Budker
Institute of Nuclear Physics, Novosibirsk 630090, Russia}
\date{\today}
\maketitle
\begin{abstract}
An eikonal model has been used to assess the relationship between
calculated strengths for first forbidden $\beta $ decay and
calculated cross sections for ($p,n$) charge exchange reactions.
It is found that these are proportional for strong transitions,
suggesting that hadronic charge exchange reactions may be useful
in determining the spin-dipole matrix elements for astrophysically
interesting leptonic processes.
\end{abstract}
 \maketitle

\section{Introduction}

Charge exchange reactions ($A_{Z},A_{Z\pm 1}$) induced by hadronic
projectiles are a powerful tool for probing spin-isospin degrees
of freedom in nuclei \cite{Os,Rap,Alf,Ej}.  The spin-isospin parts
of the operators that mediate charge exchange reactions such as
($p,n$) are the same as those involved in the corresponding
processes induced by electromagnetic and weak interactions. As a
result, the matrix elements that describe hadronic charge exchange
reactions are closely related to those that describe the rates of
$\beta$ decay or the cross sections of reactions induced by
neutrinos. It would be fortunate if this relationship were
quantitatively accurate, since it is often difficult to study the
leptonic processes directly. For example, the range of excitation
energy kinematically accessible in a $\beta$ decay transition does
not encompass the majority of the allowed (Gamow-Teller) strength
and the experimental study of neutrino induced reactions is
difficult.

A promising direction of future activity is to determine leptonic
strengths for otherwise inaccessible nuclides by studying charge
exchange reactions using radioactive (secondary) beams in inverse
kinematics \cite{JB}. This would provide nuclear properties
important for problems of nuclear physics, particle physics,
astrophysics and cosmology \cite{Ej,kosh,kubo}. One could clarify
the relationship between the spatial properties of nuclear halo
systems and the nature of soft multipole modes \cite{Daito}. One
could also determine the strength of neutrino-nucleus interactions
needed to describe the chemical evolution of the Universe,
especially the abundances of the light elements \cite{Woosley90}
and the products of $r$-process nucleosynthesis \cite{haxt}, and
to calibrate terrestrial detectors of supernova
neutrinos~\cite{full}.

However, it is not obvious a priori that the correlation of charge
exchange and leptonic matrix elements is sufficiently close for
this purpose. In contrast to leptonic processes, hadronic
reactions involve operators that have an additional radial
dependence and are subject to distortion by complex nuclear
potentials; the medium renormalization of effective operators and
the contributions of multi-step processes introduce additional
uncertainties. Therefore it was an important advance to establish
that there is an approximate proportionality between the cross
section of charge exchange reactions at very forward angles
leading to the Gamow-Teller (GT) excitations (with transferred
$T=1,\,L=0,\,J=S=1$) and the transition strength $B$(GT)
determined by intrinsic nuclear matrix elements
\cite{Daito,Good,Tadd,Xu,Fuji,Naka}. The proportionality has been
confirmed for strong transitions in a variety of nucleon and
nucleus induced charge exchange reactions; a more detailed
analysis is needed for weak GT processes~\cite{Sam}. The
model-independent character of the relation between the $L=0$
cross section for reactions induced by $^{12}$C projectiles and
the GT strength was clarified by a theoretical analysis
\cite{ostGT} based on a sensitivity function which identified the
important part of the target transition density in momentum space.

In contrast to GT transitions, the first forbidden matrix elements
of weak processes explicitly include orbital degrees of freedom.
The corresponding nuclear response in the $L=1$ channel is
associated with the states forming the spin-dipole and
giant-dipole charge-exchange resonances (SDR and GDR). This
excitation was discovered \cite{Bai} and studied on different
targets \cite{Carl} mostly using ($p,n$) reactions. Recently the
energy splitting of the $L=1$ charge exchange resonances (GDR and
SDR) was determined \cite{sam1}. There is very little information
about the quantitative relationship of charge exchange cross
sections and leptonic strength for these excitations. Here we take
a first step in providing this information by studying the
relationship between first forbidden strength and ($p,n$) cross
sections, both calculated from the same wave functions. We show
that, at the same level of accuracy as for the GT case, for strong
transitions one can expect an approximate proportionality between
the observed cross sections of charge exchange reactions
populating spin-dipole states and the corresponding nuclear
transition probabilities. We follow the general approach that was
successfully applied to  GT excitations by Osterfeld {\sl et al.}
\cite{ostGT}, extended to describe $L=1$ transitions and the
effects of the real part of the optical potential.

\section{Theoretical framework}

 In order to understand the relationship between the charge exchange
cross section and the nuclear response strength, we need to
examine the effects specific to the excitation of the SDR. For
this purpose we consider the influence of distortion in a simple
eikonal approximation (EA). In many cases, even at rather low
energy, the EA gives a good qualitative description of the
reaction cross section in the SDR region.  We obtain the relevant
wave functions and transition densities from the shell model.

The strength $B_J$ of the SDR in the long wavelength limit and the
transition form-factors $F_{J}(q)$ are calculated in terms of the
elements of the single-particle density matrix of the given
transition $i\rightarrow f$ from the ground state,
$\rho_{fi}(\nu,\nu')=\langle f|a^{\dagger}_{\nu'}a_{\nu}|
i\rangle$,
\begin{equation}
B_{J}(i\rightarrow f)=\left|\sum_{\nu\nu'}\rho_{fi}
(\nu,\nu')(\nu'||r\,O_J||\nu)\right|^2,
\label{0}
\end{equation}
\begin{equation}
F_J(q)
=\sum_{\nu\nu'}\rho_{fi}(\nu,\nu')(\nu'||j_1(qr)\,O_J||\nu),
                                                                 \label{e1}
\end{equation}
where $(\nu'||j_1(qr)\,O_J||\nu)$ is the product of the radial
matrix element of the spherical Bessel function and the reduced
matrix element of the charge exchange spin-dipole operator
$O_{JM}=\{\sigma \otimes Y_{1}({\bf \hat{r}})\}_{JM}\tau^{\pm}$.
In the limit of low momentum transfer, $qR \ll 1$, the squared
transition form-factor (\ref{e1}) is related directly to the
strength of the SDR,
\begin{equation}
F_J^2(q)\rightarrow \frac{q^2}{9}B_{J}(i\rightarrow f).
\label{1}
\end{equation}

In the case of the GT resonance a direct proportionality between
the experimentally measured cross sections of charge exchange
reactions at very forward angles and nuclear matrix elements was
confirmed by a number of studies
\cite{Daito,Good,Tadd,Xu,Fuji,Naka} at a level of accuracy of
10-15\%. From the viewpoint of the underlying physics, this
important result is based on a single-step mechanism for the
process, a simple bare operator which does not include orbital
degrees of freedom, and the dominance of the central spin-isospin
interaction $V_{\sigma\tau}$ over a broad range of energies. It is
a priori unclear whether these features pertain to the SDR case
($\Delta L=1$). At the maximum of the differential cross section
for the SDR, $qR\sim 1$. The $q$-dependence of the hadronic
operator and the effects of tensor forces may show up at the
larger $q$. As a result, there may not be a simple relationship
between the cross section and the nuclear response strength.

In the SDR case, the direct part of the $(p,n)$ reaction amplitude
(the exchange part will be discussed later) is
\begin{equation}
T_{fi}^{dir} =\int\,d^3{\bf r}\,\chi_f^{(-)\ast}({\bf k}_f,{\bf
r}) \int\,\frac{d^3{\bf q}'}{(2\pi)^3}F_J(q')V_{JM}({\bf
q}')\exp{(-\imath {\bf q}'\cdot{\bf r})}\chi_i^{(+)}({\bf
k}_i,{\bf r}),             \label{2}
\end{equation}
where the effective operators for the channels with angular
momenta $J^{\pi}=0^{-},1^{-}$ and $2^{-}$ contain contributions
from central and tensor forces,
\begin{equation}
V_{JM}({\bf q}')=V^{c}_{JM}({\bf q}')+V^{t}_{JM}({\bf q}'),
\label{3}
\end{equation}
\begin{equation}
V^{c}_{JM}({\bf
q}')=4\pi\imath\sqrt{\frac{2}{2J+1}}t^c_{\sigma\tau}(q')
\{\sigma\otimes Y^{\ast}_1(\hat{\bf q}')\}_{JM},
\label{4}
\end{equation}
\begin{equation}
V^{t}_{JM}({\bf q}')=4\pi\imath\sqrt{\frac{2}{2J+1}}t^t_\tau(q')
[\{\sigma\otimes Y^{\ast}_1(\hat{\bf q}')\}_{JM}
-3(\mbox{\boldmath $\sigma$}\cdot \hat{\bf q}')\{\hat{\bf
q}'\otimes Y^{\ast}_1(\hat{\bf q}')\}_{JM}].
\label{e2}
\end{equation}
In Eqs. (\ref{4}) and (\ref{e2}), $t^c_{\sigma\tau}(q')$ and
$t^t_\tau(q')$ are, respectively, the central and tensor
components of the nucleon-nucleon $t$-matrix, see Franey and Love
\cite{LF85}, that are responsible for spin-isospin transfer. The
excitation of different $J$-components of the SDR proceeds via
different combinations of the amplitudes of the nucleon-nucleon
effective interaction. For the $0^-$ part, the tensor interaction
can be combined with the central one using the relations
\begin{equation}
\{\hat{\bf q}'\otimes Y^{\ast}_1(\hat{\bf
q}')\}_{00}=-\frac{1}{\sqrt{4\pi}}, \quad \{\sigma\otimes
Y^{\ast}_1(\hat{\bf q}')\}_{00}=- \frac{(\mbox{\boldmath
$\sigma$}\cdot \hat{\bf q}')}{\sqrt{4\pi}}. \label{7}
\end{equation}
These identities produce the combination
$t^{l}(q)=t^c_{\sigma\tau}(q)- 2t^t_\tau(q)$ which is just the
spin-longitudinal component of the nucleon-nucleon $t$-matrix. For
the $1^-$ part, the operator $\{\hat{\bf q}' \otimes
Y^*_1(\hat{\bf q}')\}_{1M}=0$, and the amplitude in Eq. (\ref{2})
becomes proportional to
$t^{tr}(q)=t^c_{\sigma\tau}(q)+t^t_\tau(q)$, which is the
spin-transverse component of the nucleon-nucleon $t$-matrix. For
the $2^-$ part, both components contribute, and the amplitude in
Eq. (\ref{2}) has a more complicated form.

The functions $\chi^{(-)}$ and $\chi^{(+)}$ in the amplitude
(\ref{2}) are optical-model wave functions describing the motion
of the initial proton and final neutron in the optical potential
of the target nucleus. To disentangle the nuclear transition
form-factor from the observed cross section, one needs to unravel
the intrinsic nuclear dynamics masked by the distorted waves
$\chi^{(\pm)}$.

\section{Distortion factor}

The effective operators (\ref{4}) and (\ref{e2}) in the reaction
amplitude (\ref{2}) are evaluated at the value ${\bf q}'$ of the
local momentum transfer that corresponds to the charge exchange
event. However, because of the distortion by the optical
potential, ${\bf q}'$ does not coincide with the asymptotic
momentum transfer ${\bf q} ={\bf k}_i - {\bf k}_f$. In the absence
of distortion (the plane wave approximation) we would have
\begin{equation}
[\chi_f^{(-)\ast}({\bf k}_f,{\bf r})\chi_i^{(+)}({\bf k}_i,{\bf
r})]_{PW}= \exp{(\imath{\bf q}\cdot{\bf r})}
\label{5}
\end{equation}
so that the integration over ${\bf r}$ could be performed
explicitly resulting in $\delta({\bf q}-{\bf q}')$. For distorted
waves this is no longer true.

In the EA, the product of two optical-model wave functions in Eq.
(\ref{5}) can be estimated by
\begin{equation}
\chi_f^{(-)\ast}({\bf k}_f,{\bf r})\chi_i^{(+)}({\bf k}_i,{\bf
r})= \exp{(\imath{\bf q}\cdot{\bf r})}D({\bf r}_\perp ),
\label{e3}
\end{equation}
where the distortion factor $D({\bf r}_\perp )$ is defined by
\begin{equation}
D({\bf r}_\perp )
=\exp{\left[-\frac{\imath}{\hbar}\int^\infty_{-\infty}
\frac{dz}{v} U_{opt}(z,{\bf r}_\perp)\right]}.
\label{e4}
\end{equation}
In the spirit of the eikonal approximation, the longitudinal
momentum is still preserved, while the distortion is effective in
the plane perpendicular to the trajectory. In Eq. (\ref{e4}), the
optical potential is different in the initial and final channels.
In first order we account for this difference by assuming a fast
single-step process which leads to
\begin{equation}
\frac{U_{opt}}{v} = \frac{1}{2}\left(\frac{U_{opt}^i} {v_i} +
\frac{U_{opt}^f}{v_f}\right).
\label{6}
\end{equation}
For a square well potential of depth $U_0$ with a sharp boundary
at $r=R$ the distortion factor can be calculated analytically:
\begin{equation}
D(r) = \exp{\left(-\imath \frac{2U_0}{\hbar v}\sqrt{R^2 -
r^2}\right)},
                                                                  \label{e5}
\end{equation}
and $D(r)=1$ for $r > R$. This approximation might be insufficient
in the region of minimum of the cross section where the details of
the potential shape are essential. However, near the maximum,
which is our region of interest, the cross section is insensitive
to the diffuseness of the optical-model potential. We have checked
this point by varying the diffuseness in a DWIA calculation.

The exchange part of the reaction amplitude was estimated in the
``fixed ${\bf Q}$'' approximation \cite{KMT59,LF81}. This
approximation has been found \cite{AM,PMAM} to improve with
increasing energy and multipolarity of the excitation for
even-state forces and mainly $S=0$ natural parity transitions. We
need to deal with  $S=1$ transitions and include tensor forces.
Recently, the fixed ${\bf Q}$ approximation for exchange was
discussed in detail in a review article \cite{baker97}; it is more
accurate for central interactions than for the tensor interaction
so that the latter is of most concern. Fortunately, the
approximation for  tensor forces is best when $Q$ is not far from
$k_{i}$, which is the case for the relatively high energies
considered here ($k_{i}$ is greater than the Fermi momentum in the
nucleus) and for small angles. It is also relatively less
important for the transitions we consider where there are large
central and direct tensor contributions \cite{love01}. Baker {\sl
et al.} \cite{baker97} note that for the intermediate energy
($p,p'$) isovector excitation of $1^{-}$ and $2^{-}$ states in
$^{40}$Ca, the cross sections obtained with the fixed ${\bf Q}$
approximation were usually within 30\% of the results of
calculations done with exchange treated exactly. As shown in Fig.
1, our eikonal calculations with the fixed ${\bf Q}$ approximation
are within 20\% of DWBA calculations with exact exchange,
consistent with the ($p,p'$) result. This level of agreement seems
adequate for our purposes: the overall effect of the approximation
is simply to change slightly the scale of the cross sections for a
given $J$ from that shown in Fig. 2. If the effect were different
for different transitions, it would show up in the overall scatter
of the points about the average line which is small for strong
transitions.

In the fixed ${\bf Q}$ approximation \cite{KMT59} the exchange
momentum ${\bf Q}$ coincides in the laboratory frame with the
initial momentum ${\bf k}_i$. For the exchange amplitude we then
obtain
\begin{equation}
T^{ex}_{fi}=\sqrt{2}\left[(\tilde{t}^c_{\sigma\tau}(k_i)+\tilde{t}^t_\tau(k_i))
\mbox{\boldmath $\sigma$} -3\tilde{t}^t_\tau(k_i)(\mbox{\boldmath
$\sigma$} \cdot \hat{\bf k}_i)\hat{\bf k}_i\right]\cdot\langle
f|{\bf O}^{ex}({\bf q})| i\rangle,
\label{e8}
\end{equation}
where the central $\tilde{t}^c_{\sigma\tau}$, and tensor,
$\tilde{t}^t_\tau$, interactions are defined by Franey and Love
\cite{LF85}. In Eq. (\ref{e8}) the effective exchange operator
\begin{equation}
{\bf O}^{ex}({\bf q})=\sum_j\exp{(\imath {\bf q}\cdot {\bf
r}_j)}D({\bf r}_{\perp j})\mbox{\boldmath $\sigma$}_j\tau^-_j
\label{9}
\end{equation}
includes the distortion factors $D(r)$ specific for each nucleon
inside the nuclear matrix element.

\section{Sensitivity function}

In Eq. (\ref{2}) the integration over ${\bf r}$ is not well
defined at large distances. It is convenient to single out a
no-distortion contribution proportional to $\delta({\bf q}-{\bf
q}')$ by using the decomposition $D(r) \rightarrow 1 +[D(r)-1]$.
The first term describes the plane wave contribution and the
second the effects of distortion. Since $\left|D(r)\right| \leq
1$, the distortion term reduces the plane wave contribution. A
convenient form of Eq. (\ref{2}) can be obtained for transitions
to $0^-$ and $1^-$ states by writing it as
\begin{equation}
T_{fi}^{dir} =\frac{\{\sigma \otimes {\rm
T}^{(J)}\}_{JM}}{\sqrt{2J+1}},
                                                                  \label{8}
\end{equation}
where
\begin{equation}
{\rm T}_m^{(J)} ={\rm T}_m^{J}(PW) + \int_0^\infty
dq'\,S_m^J(q,q')F_J(q')
                                                                \label{e6}
\end{equation}
is the amplitude describing the excitation of the SDR with the
longitudinal, $m=0$, or transverse, $m=\pm 1$, relative
proton-neutron spatial oscillations. In Eq. (\ref{e6}), ${\rm
T}_m^{(J)}(PW)$ is the plane wave contribution, and we have
introduced the sensitivity function \cite{ostGT}
$$
S_m^J(q,q') =\frac{2\sqrt{2}}{\pi}q'^2\int d^{3}{\bf
r}\,\exp{(\imath {\bf q}\cdot{\bf r})}(D({\bf
r}_\perp)-1)Y_{1m}(\hat{\bf r})j_1(q'r)
$$
\begin{equation}
\times \left\{ \begin{array}c t^l(q')\;{\rm for}\; J=0, \\
                         t^{tr}(q')\;{\rm for}\;
J=1\end{array}. \right.
\label{e7}
\end{equation}
The sensitivity function in Eq. (\ref{e7}) characterizes the range
of $q'$ which contribute importantly to the charge exchange cross
section for a given asymptotic momentum transfer $q$.

\section{Example: $^{12}$C$({\protect \small p,n})^{12}$N reaction}

As an example of application of the method we performed numerical
calculations for the $^{12}$C$(p,n)^{12}$N reaction to compare
with experimental data \cite{ACM96} for the excitation of
spin-dipole states at a proton energy of 135 MeV.

For $^{12}$C, with the optical-model potential of Ref.
\cite{CK80}, $D(r)$ varies smoothly inside the nucleus. Near the
surface it changes rapidly from its value at the center
$D(0)\approx 0.5$, to the value of 1. It is then a good
approximation to write the exchange matrix element in Eq.
(\ref{e8}) as
\begin{equation}
\langle f|{\bf O}^{ex}({\bf q})|i\rangle \approx D(r_0)\langle
f|\sum_j \exp{(\imath {\bf q}\cdot{\bf r}_j)}\mbox{\boldmath
$\sigma$}_j\tau^-_j|i \rangle.
\label{e9}
\end{equation}
The result is not very sensitive to a particular choice of the
reference point $r_0$; we used $r_0=0$. This approximation
underestimates the exchange part. Near the maximum of the cross
section its contribution is not significant. It becomes important
at large angles where the difference of distortion along different
trajectories is noticeable; however, the cross section at large
angles is small. We mention parenthetically that, opposite to the
results of \cite{AM,PMAM} for even-state forces, the exchange
contribution reduces the cross section. This is related to our
inclusion of odd-state forces whose net effect is small after
exchange is accounted for.

In our calculations, the wave functions and transition densities
for the spin-dipole states were obtained using a harmonic
oscillator basis including the orbitals of $p,\, sd$ and $pf$
shells that form the $3\,\hbar \omega$ model space necessary for
the description of the $L=1$ excitations. The calculations were
performed with the WBN residual interaction and a harmonic
oscillator parameter of $1.64(A/A-1)^{1/2}$  fm \cite{B}. The
cross sections for the $^{12}$C($p,n$)$^{12}$N reaction leading to
the $1^-$ state at $E_{x}=1.8\,$MeV and the $2^-$ state at
$E_x=4.3$ MeV were calculated as the sum of the direct and
exchange amplitudes, Eqs. (\ref{e2}) and (\ref{e8}). The results
are shown in Fig. 1 together with data from Ref. \cite{ACM96}. For
comparison, a calculation with the DW81 code is also presented.
The calculations give similar cross section shapes near the
maximum and  significantly overestimate the magnitude of the cross
section. The results are very similar for other excited states.
They are also similar to the distorted wave results obtained for
the same transition in Ref. \cite{ACM96} using a $1\,\hbar\omega$
model space and the MK interaction.

The systematics of the cross sections at their maximum divided by
the calculated $\beta$ decay strengths are shown in Fig. 2 for
$0^-,\,1^-$ and $2^-$ states. As seen from Fig. 2, there is an
approximate proportionality between the cross section at the
maximum and the spin-dipole strength, accurate to within 10-15\%,
for states with strength $B_{J}>0.1\;{\rm fm}^2$. This is the same
level of proportionality as for GT ($L=0$) excitations at very
forward angles. One may ask whether the validity of this
conclusion is affected by the poor agreement in the magnitude of
the cross section for the 1.8 MeV $1^{-}$ state. We would argue
that this is not the case: since the wave functions are
sufficiently complex, they provide a reasonable sample of possible
behavior with respect to the operators involved. Furthermore, one
might expect proportionality to fail for such weak transitions.

\section{Discussion}

It is not clear a priori that the high degree of proportionality
shown in Fig. 2 should occur. The cross section involves an
integral of the transition form factor over a range of $q'$ while
the value of $B_J$ is determined by evaluating the form factor at
$q'\approx 0$. To examine what leads to the observed
proportionality, we return to Eq. (\ref{e6}). Two factors
determine the ($p,n$) cross section: the transition form-factor
$F_J(q)$ and the sensitivity function $S^J_m(q,q')$. In Fig. 3 we
show the transition form-factors for different $1^-$ states
normalized to the same maximum value in order to compare their
shapes. The shapes are very similar near the maximum but differ at
higher momentum transfer $q$. If the region of high $q'$ does not
contribute significantly in the integration over $q'$ in Eq.
(\ref{e6}), the integrals for different form-factors will be
proportional.

Samples of the imaginary parts of the  sensitivity functions are
shown in Figs. 4 and 5;  the real parts have very similar shapes
and are typically a factor of two smaller in magnitude. A general
remark should be made about the $q'$-dependence at small $q'$.
Since $D({\bf r}_\perp)$ does not depend on the longitudinal
coordinate $z$, the $z$-component of the local momentum transfer
${\bf q}'$ must coincide with the $z$-component of the asymptotic
momentum transfer ${\bf q}$. When the absolute value of $q'$ is
smaller than $q_z$, this condition cannot be fulfilled at any
angles of ${\bf q}'$, and the sensitivity function must be equal
to zero.

As noted above, the sensitivity function for $J=0$ is proportional
to the spin-longitudinal component of the effective interaction,
and that for $J=1$ to the spin-transverse one. We, therefore,
expect a different $q'$-dependence reflecting the different
behavior of $t^l(q')$ and $t^{tr}(q')$. For $0^-$ states the
projection $m=0$ dominates, corresponding to the spin-longitudinal
behavior of the reaction amplitude for $J=0$; the sensitivity
function for $m=1$ is smaller by an order of magnitude. At the
small scattering angle corresponding to the peak of the cross
section, $\theta =4.3^{\circ}$, the momentum transfer ${\bf q}$ is
almost parallel to the initial proton momentum ${\bf k}_i$, thus
enhancing the $m=0$ component. For $J=1$ the picture is different,
as is seen in Fig. 5. Projections $m=0$ and $m=\pm 1$ give
comparable contributions.

Given the nature of the sensitivity functions it is clear why the
cross sections and $B_J$ are closely proportional. For both $J=0$
and $J=1$ the main contribution comes from the peak region where
the transition form factors have the same shape, leading to the
observed proportionality.

In summary, our results imply that there will be an approximate
proportionality of the observed cross section at the maximum of
the charge exchange reaction  exciting spin-dipole modes and the
leptonic strength. This supports the possibility of using such
reactions for extracting leptonic strengths of astrophysical
interest.  Having established here the basic apparatus to examine
this issue, it will next be important to consider transition
densities for heavier nuclides, so as to determine whether their
shapes are similar enough that cross sections and $B_J$ strengths
will be proportional. It will also be important to examine the
nature of the sensitivity functions for heavier nuclei, to
ascertain whether they remain concentrated in a relatively small
range of $q'$ where the transition densities are similar.
\begin{center}
ACKNOWLEDGMENTS
\end{center}
The authors are thankful to W. G. Love for a helpful discussion.
The shell model information necessary for the calculations was
kindly provided by B. A. Brown. V.F.D. is thankful to T. Nakamura
for assistance and discussions and to the NSCL for the hospitality
and support. This work was supported by the National Science
Foundation.

\newpage

\begin{figure}
\caption{Cross sections for the reaction $^{12}$C($p,n$)$^{12}$N
leading to the $1^-$ state at $E_x=1.8$ MeV and the $2^-$ state at
$E_x=4.3$ MeV. The cross sections shown as solid lines are the
results of the eikonal approximation calculations described here;
a DWIA calculation done with the same parameters is shown by solid
dots. The data shown as solid squares are from Ref. \protect
\cite{ACM96}, as are the DWIA calculations shown as open circles.
All the theoretical calculations have been multiplied by the
factor shown in the Figure.} \label{Fig. 1.}
\end{figure}

\begin{figure}
\caption{Ratios of the cross sections for excitation of the
spin-dipole states, taken at their maximums, to the corresponding
spin-dipole strengths $B_J$ for states with different $B_J$. The
upper panel is for $0^-$ states; the middle panel for $1^-$
states; and the lower panel for $2^-$ states.} \label{Fig. 2.}
\end{figure}

\begin{figure}
\caption{Transition form-factors for the $1^-$ states normalized
to their maximum values. The state with the anomalous shape
corresponds to the high point near $B_1 =0.45\,$fm$^2$ in Fig. 2,
middle panel.} \label{Fig. 3}
\end{figure}

\begin{figure}
\caption{The imaginary part of the sensitivity functions
$S_{m}(q,q')$ for $J=0$. The small size of $S_1$ reflects the
spin-longitudinal origin of the reaction amplitude.}
\label{Fig. 4.}
\end{figure}

\begin{figure}
\caption{The imaginary part of the sensitivity functions
$S_{m}(q,q')$ for $J=1$. $S_1$ and $S_0$ are comparable for this
spin-transverse dominated reaction amplitude.}
\label{Fig. 5.}
\end{figure}

\end{document}